\documentclass[letter, twocolumn]{jpsj3}
\usepackage{txfonts}
\usepackage{braket}
\usepackage{bm}

\usepackage{graphicx}
\usepackage{dcolumn}
\usepackage{bm}
\usepackage{amsmath}
\usepackage{times}
\usepackage[usenames]{color}

\newcommand{\kets}[1]{|#1\rangle}

\newcommand{\means}[1]{\langle#1\rangle}

\def\journal #1#2#3#4{#1 {\bf #2}, #3 (#4)}
\def\PR{Phys.\ Rev.}

\def\PRB{Phys.\ Rev.\ B}

\def\PRL{Phys.\ Rev.\ Lett.}

\def\JPSJ{J.\ Phys.\ Soc.\ Jpn.}

\def\RMP{Rev.\ Mod.\ Phys.}

\def\JACS{J.\ Am.\ Chem.\ Soc.}

\title{Magnetic field effects in a correlated electron system with spin-state degree of freedom \\
- Implication of an excitonic insulator -}
\author{Taro Tatsuno$^{1}$, Eriko Mizoguchi$^{1}$, Joji Nasu$^{2}$, Makoto Naka$^{1}$  and Sumio  Ishihara$^{1}$\thanks{ishihara@cmpt.phys.tohoku.ac.jp}}
\inst{$^1$Department of Physics, Tohoku University, Sendai 980-8578, Japan \\
$^2$  Department of Physics, Tokyo Institute of Technology, Meguro, Tokyo 152- 8551, Japan
} 

\abst{
Magnetic field $(H)$ effects on a correlated electron system with the spin-state degree of freedom are examined. 
The effective Hamiltonian derived from the two-orbital Hubbard model is analyzed by the mean-field approximation. 
Applying $H$ to the low-spin (LS) phase induces the excitonic insulating phase, as well as the spin-state ordered phase where the LS and high-spin (HS) states are ordered alternately. 
In the case where $H$ is applied to the HS phase, a reentrant transition for the HS phase appears. 
A rich variety of the phase diagrams are attributed to the spin-state degree of freedom and their combinations in the wave function as well as in the real-space configuration. 
Present results provide a possible interpretation for the recent experimental observation under the strong magnetic field in LaCoO$_3$. 
}

\kword{spin state, magnetic field, excitonic insulator}

\begin{document}
\maketitle


Spin-state degree of freedom (SSDF) in the transition-metal ions have been widely recognized as one of the indispensable factors which provide rich variety of physics in magnetic solids. 
There are multiple spin states in a single magnetic ion due to the different local electron configurations. 
Transitions between the multiple spin states are often seen in the iron and cobalt ions which are included not only in correlated electron materials~\cite{Imada1998,tokura1998,Gretarsson2013}, but also in biomaterials,~\cite{Halder2002} and earth inner-core materials~\cite{Lin2005,Antonangeli2011,Hsu2011}. 
A driving force of the spin-state transition is attributed to a competition between the crystalline-field effect and the Hund's coupling, which stabilize the low-spin (LS) and high-spin (HS) states, respectively.~\cite{Werner2007,suzuki}

Perovskite cobalt oxides and their derivatives are the prototypical examples of the correlated electron materials with SSDF. 
A nominal valence of the cobalt ion in LaCoO$_3$ is 3+ where the number of electrons occupying the $3d$ orbitals is six. 
The characteristic temperature dependences of the electrical resistivity and the magnetic susceptibility are interpreted as a crossover between the LS state of the $(t_{2g})^6$ configuration 
with $S=0$ and the HS states of $(t_{2g})^4(e_g)^2$ with $S=2$.~\cite{tokura1998}
Several exotic phenomena, such as the giant magnetoresistance,~\cite{Mahendiran1996} magnetic clusters,~\cite{sato,yu2009} and a ferroelectricity~\cite{oka2010}, are attributable to the spin-state change. 

Correlated electron materials with SSDF is recently reexamined as a plausible candidate of the excitonic insulator (EI).~\cite{kunes1,kunes2,nasu} 
The EI state has been studied since 1960s in the narrow gap semiconductors and semimetals~\cite{mott1961,jerome1967,halperin1968,Fukuyama1971,Kuramoto1978}, and is recently reexamined from the modern viewpoints.~\cite{wakisaka,kaneko}
The EI state is expected to be realized, when the exciton binding energy exceeds the band gap energy. 
Since the LS and HS states in the cobalt oxides are identified as a band insulator and a Mott insulator, respectively, 
a narrow-gapped state is expected around a crossover between the two spin states. 
Actually, a phase transition around 90K observed in Pr$_{0.5}$Ca$_{0.5}$CoO$_3$ is examined 
from the viewpoints of the EI phase transition.~\cite{{kunes2}}

It is required to develop how to control the spin states as well as the EI state in correlated electron materials with SSDF. 
It is well known that applying pressure stabilizes the LS state relative to the HS state owing to its small ionic radius~\cite{Vanko2006,oka2010}.
The ion substitution, the thin-film synthesis, and the pulse-laser irradiation also bring about the spin-state change. ~\cite{Fujioka2015,kanamori1,kanamori2,Okimoto2013,Ishikawa2013}.
Magnetic field is a promising tool to control the spin states. 
The experimental and theoretical investigations for the magnetic-field induced spin-state change and the EI phase have recently  
attracted much attention~\cite{altrawneh,rotter,platonov,ikeda,kunes3}.  
This is expected as a powerful and simple tool in comparison with other techniques. 

In this Letter, the magnetic field effects in correlated electron systems with SSDF are studied. 
We analyze the effective Hamiltonian derived from the two-orbital Hubbard model under the magnetic field, in which the LS and HS states are adopted as the basis states. 
The phase diagrams under the magnetic field are obtained by the mean-field (MF) approximation. 
When the magnetic field is applied to the LS phase, the EI phase is stabilized owing to the magnetic moment due to the HS component in the wave function. 
By further increasing the magnetic field, sequential phase transitions occur to the LS-HS ordered phase. 
In the case where the magnetic field is applied to the HS phase, 
a reentrant transition for the HS phase occurs owing to the two-sublattice structure for SSDF. 
A possible interpretation of the recently observed magnetic-field induced phase is proposed.

We start from the two-orbital Hubbard model with a finite-energy difference between the orbitals defined by 
$ {\cal H}_{\rm TH}={\cal H}_t+{\cal H}_u$. 
Each term is defined by 
\begin{align}
{\cal H}_t=-\sum_{\left < ij \right > \eta\sigma}
t_\eta 
(c_{i\eta \sigma}^\dagger c_{j\eta \sigma}+{\rm H.c.}) , 
\label{eq:Ht}
\end{align}
and 
\begin{align}
 {\cal H}_u&=\Delta\sum_i n_{i a} +U\sum_{i \eta}n_{i\eta\uparrow}n_{i\eta\downarrow}+U'\sum_{i} n_{ia}n_{ib}\nonumber\\
&+J\sum_{i\sigma\sigma'}c_{ia\sigma}^\dagger c_{ib\sigma'}^\dagger c_{ia\sigma'}c_{ib\sigma}
 +I\sum_{i\eta\neq \eta'}c_{i\eta\uparrow}^\dagger c_{i\eta \downarrow}^\dagger c_{i\eta'\downarrow}c_{i\eta'\uparrow} . 
\label{eq:Hu}
\end{align}
We define the creation (annihilation) operator for an electron with orbital $\eta(=a,b)$ and spin $\sigma (=\uparrow,\downarrow)$ at site $i$ as $c_{i\eta \sigma}^\dagger$ ($c_{i\eta \sigma}$). 
The number operator is defined as $n_{i\eta}=\sum_\sigma n_{i\eta\sigma}$ 
with $n_{i\eta\sigma}=c_{i\eta\sigma}^\dagger c_{i\eta\sigma}$.  
Equation~(\ref{eq:Ht}) describes the electron hopping 
with amplitude $t_\eta$ between the same orbital $\eta$ in the nearest neighbor (NN) sites. 
The first term in Eq.~(\ref{eq:Hu}) represents an energy difference, $\Delta (>0)$, between the two orbitals.  
The other terms represent the on-site electron-electron interactions where $U$, $U'$, $J$, and $I$ are the intra-orbital Coulomb interaction, the inter-orbital Coulomb interaction, the Hund coupling, and the pair-hopping interaction, respectively.
We fix an average electron number per site at two.   

%
In order to examine the competition between the multiple spin states,
we derive the effective Hamiltonian for the low-energy electronic states, where ${\cal H}_U$ and ${\cal H}_{t}$ are  the unperturbed and perturbed terms, respectively. 
There are the six electronic states in ${\cal H}_{U}$, when the electron number at a site is fixed at two. 
Among them, 
the low-energy spin-singlet and -triplet states are taken as the basis states, and are termed the LS and HS states, respectively, in this paper. 
The wave functions are given by 
$ \kets{L}=(f c_{b\uparrow}^\dagger c_{b\downarrow}^\dagger - g c_{a\uparrow}^\dagger c_{a\downarrow}^\dagger )\kets{0}$, 
and 
$\{ \kets{H_{+1}},  \kets{H_{0}}, \kets{H_{-1}} \}$
with 
$ \kets{H_{+1}}=c_{a\uparrow}^\dagger c_{b\uparrow}^\dagger\kets{0}$, 
$ \kets{H_{0}}=1/\sqrt{2} (c_{a\uparrow}^\dagger c_{b\downarrow}^\dagger + c_{a\downarrow}^\dagger c_{b\uparrow}^\dagger )\kets{0}$, and 
$ \kets{H_{-1}}=c_{a\downarrow}^\dagger c_{b\downarrow}^\dagger\kets{0}$. 
We introduce $f=1/\sqrt{1+(\Delta'-\Delta)^2/I^2}$ and $g=\sqrt{1-f^2}$.~\cite{kanamori1,kanamori2,nasu}
%
The low-energy effective Hamiltonian for the two-orbital Hubbard model is obtained 
by the perturbational processes. 
Details are presented in Ref.~\citen{nasu}. 
We show the results as 
\begin{align}
 {\cal H}&= -h_z\sum_i\tau_i^z+J_{z}\sum_{\means{ij}}\tau_i^z\tau_j^z
+J_s\sum_{\means{ij}}\bm{S}_i\cdot\bm{S}_j\nonumber\\
&-\sum_{\means{ij}} \sum_{\Gamma=(0, \pm 1)}
\left ( J_x \tau_{\Gamma i}^x\tau_{\Gamma j}^x+J_y\tau_{\Gamma i}^y\tau_{\Gamma j}^y \right )  
 , 
\label{eq:Heff}
\end{align}
where ${\bm S}_i$ is the spin operator with amplitude of $S=1$. 
We introduce the operators for SSDF defined by  
$\tau_\Gamma^{x}=\kets{L}\bra{H_\Gamma}+\kets{H_\Gamma}\bra{L}$, $\tau_\Gamma^{y}=i(\kets{L}\bra{H_\Gamma}-\kets{H_\Gamma}\bra{L})$,  and $\tau_\Gamma^{z}=\kets{H_\Gamma}\bra{H_\Gamma}-\kets{L}\bra{L}$ for $\Gamma=(+1, 0, -1)$, 
and 
$\tau_i^\gamma=\sum_\Gamma \tau_{ \Gamma i}^\gamma$. 
The energy parameters are given by the parameters in the two-orbital Hubbard model, 
and are defined to be positive in the present calculation.~\cite{nasu} 
The first term in Eq.~(\ref{eq:Heff}) represents the crystalline field effect, and 
the second term favors an alternate order of the LS and HS states, termed the LS/HS ordered state. 
The last two terms stabilize the EI phase, which is identified by the order parameters 
$\langle \tau^x_\Gamma \rangle$ and $\langle \tau^y_\Gamma \rangle$. 
The wave function in the EI phase is represented by 
\begin{eqnarray}
|\psi_{\rm EI} \rangle =C_L |L \rangle+\sum_{\Gamma} C_H^\Gamma| H_\Gamma \rangle ,
\label{eq:wfei}
\end{eqnarray}
with the complex numbers $C_L$ and $C_H^\Gamma$. 
In addition to the Hamiltonian in Eq.~(\ref{eq:Heff}), 
the Zeeman energy term is introduced as 
\begin{eqnarray}
{\cal H}_{zeeman}=- H  \sum_{i } S^z_{i } ,
\end{eqnarray}
where the magnetic field is parallel to the $z$ axis.

\begin{figure}[t]
\includegraphics[width=0.85\columnwidth,clip]{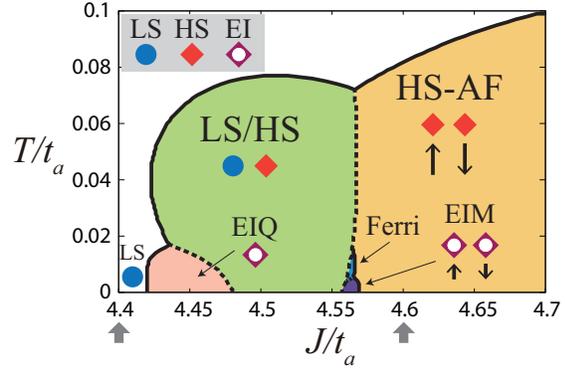}
\caption{(Color online) 
Finite $T$ phase diagram without $H$. 
Solid and broken lines represent the second- and first-order phase transitions, respectively. 
Circles, rhombuses, and their combined symbols represent the LS, HS, and EI states, respectively, 
and arrows represent schematic spin directions. 
Bold arrows denote the values of $J$ in which the numerical results shown in Figs.~\ref{fig:TH} and \ref{fig:MH2} are calculated. 
A ratio of the hopping integrals is chosen to be $t_b/t_a=-0.1$.
}
\label{fig:TJ}
\end{figure}
Electronic structure under the magnetic field is analyzed by the MF approximation. 
The two-body terms in Eq.~(\ref{eq:Heff}) are decoupled as 
$O_i O_j \rightarrow \means{O_j}O_j+O_i \means{O_j}-\means{O_i}\means{O_j}$ where 
$\means{O_i}$ represents the thermal average of the local operator. 
The order parameters $\means{\bm S}$ and $\means{\tau^\gamma_\Gamma}$ are 
considered in the two sublattices termed A and B in a square lattice. 
%
We vary mainly the Hund coupling $J$ and the magnetic field $H$, and fix other parameters at  $\Delta/t_a=8$, $U=4J$, $U'=3J$, and $I=J$ in the numerical calculations. 

First, we show the finite-temperature ($T$) phase diagram without magnetic field ($H$) in Fig.~\ref{fig:TJ}. 
As presented in Ref.~\citen{nasu}, the electronic structures at $T=0$  is changed 
with increasing $J$ as LS$ \rightarrow$EIQ$\rightarrow$LS/HS$\rightarrow$EIM$\rightarrow$HS-AF, where 
EIQ and EIM denote the two types of the EI phases, LS/HS denotes a staggered order phase of the HS and LS states, and HS-AF represents the HS phase with the antiferromagnetic (AF) order. 
Both the EIQ and EIM phases are characterized by the uniform component of the EI order parameter  $\means{\tau^x}$, and 
are distinguished by the magnetic structures; whereas the AF order is realized in the EIM phase, the spin-quadrupole order characterized by $\means{S^\alpha}=0$ and $\means{S^\alpha S^\beta } \ne 0$ 
appears in the EIQ phase. 
With increasing $T$ in Fig.~\ref{fig:TJ}, both the EI phases disappear and the LS/HS phase is stabilized in a wide region of $J$. 
This is owing to the spin entropy at the HS sites in the LS/HS phase, in which the exchange interactions between the NN HS sites do not work in the present model Hamiltonian. 
Above the EIM phase around $J/t_a=4.56$, we confirm the ferrimagnetic phase termed Ferri, in which the  spin moments appear at the LS sites in the LS/HS ordered state. 
It is worth mentioning  that this Ferri phase may correspond to the ferrimagnetic phase observed recently in the epitaxial LaCoO$_3$ sample~\cite{Fujioka2015}. The present results suggest a possibility of the EIM phase in the low temperature side of the ferrimagnetic phase. 

\begin{figure}[t]
 \includegraphics[width=0.85\columnwidth,clip]{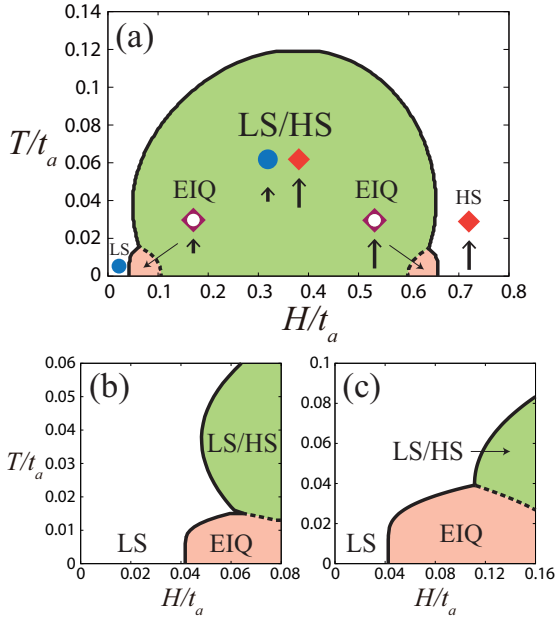}
\caption{(Color online) 
(a) Finite $T$ phase diagram under $H$ applied to the LS phase. 
Solid and broken lines represent the second- and first-order phase transitions, respectively. 
Circles, rhombuses, and their combined symbols represent the LS, HS, and EI states, respectively, 
and arrows represent schematic spin directions. 
Parameter values are chosen to be $J/t_a=4.4$, corresponding to the bold arrow in Fig.~\ref{fig:TJ}, and 
$t_b/t_a=-0.1$. 
(b, c) $H-T$ phase diagrams in the weak $H$ regions for $(J/t_a, t_b/t_a)=(4.4, -0.1)$ in (b), and for $(4.35, -0.2)$ in (c). 
}
\label{fig:TH}
\end{figure}
The finite-$T$ phase diagram under $H$ is presented in Fig.~\ref{fig:TH}(a), where 
$H$ is applied to the LS phase at $J/t_a=4.4$ (see the bold arrow in Fig.~\ref{fig:TJ}). 
In low $T$, the LS phase is changed into the EIQ phase by applying $H$, i.e. the magnetic-field induces the EI phase. 
With increasing $H$, sequential phase transitions occur as 
LS$\rightarrow$EIQ$\rightarrow$LS/HS$\rightarrow$EIQ$\rightarrow$HS. 
%
The phase diagrams under the weak magnetic field are presented in Figs.~\ref{fig:TH}(b) and \ref{fig:TH}(c). 
Relation between these results and the experimental observations is discussed later. 

\begin{figure}[t]
\includegraphics[width=0.85\columnwidth,clip]{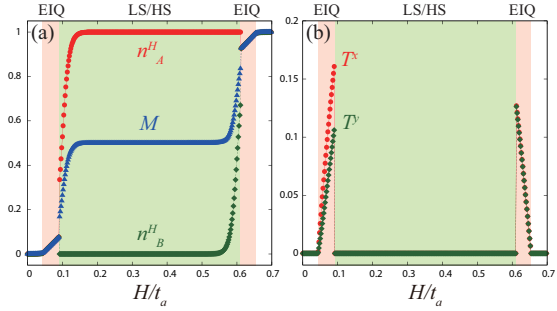}
\caption{(Color online) 
Magnetic field dependence of (a) the HS densities ($n^H_{A, B}$) and magnetization ($M$), and  (b) the squares of the EI order parameters ($T^{x, y}$). 
 Magnetic field is applied to the LS phase. 
Parameter values are chosen to be $J/t_a=4.4$ and $t_b/t_a=-0.1$. 
}
\label{fig:NH1}
\end{figure}

The $H$ dependences of the physical quantities at $J/t_a=4.4$ and $T/t_a=0.01$ are presented in Fig.~\ref{fig:NH1}. 
We define the HS density $n^{H}$ as a thermal expectation value of 
$\sum_{\Gamma}   |H_{i \Gamma} \rangle \langle H_{i \Gamma} |$, and the squares of the uniform parts of the EI order parameter 
$T^{x(y)} = (1/4)\sum_\Gamma
( \langle \tau_{\Gamma }^{x(y)}\rangle_A+\langle \tau_{\Gamma }^{x(y) }\rangle_B)^2$, 
where $\langle \tau_{\Gamma }^{x(y)}\rangle_{A, B}$ are the expectation values of $\tau_{\Gamma }^{x(y)}$ in the sublattices A and B. 
The magnetization $(M)$ does not appear in the LS phase owing to the finite spin gap. 
In the EIQ phase appearing around $H/t_a=0.04$, a small magnetization associated with the uniform components of the EI order parameters appear.  
The magnetic moment in this phase is attributed to the HS component in the EI wave function in Eq.~(\ref{eq:wfei}). 
The first-order phase transition from the EIQ phase to the LS/HS phase occurs around $H/t_a=0.09$, 
and the HS density in the sublattice A ($n^H_{A}$) increases and that in the sublattice B($n^H_{B}$) decreases.
The Zeeman energy is gained in the HS sublattice A, while almost non-magnetic LS state is realized in the sublattice B. 
A magnetization plateau with $M=1/2$ emerges in this region. 
At around $H/t_a=0.6$, the reentrant transition to the EIQ phase occurs. 
With increasing $H$ near the phase boundary in the LS/HS phase, $n^H_{A}$ decreases and $n^H_{B}$ rapidly increases,  and the two-sublattice structure for SSDF disappears.  
With further increasing $H$, the HS state is realized in all sites.  

\begin{figure}[t]
\includegraphics[width=0.85\columnwidth,clip]{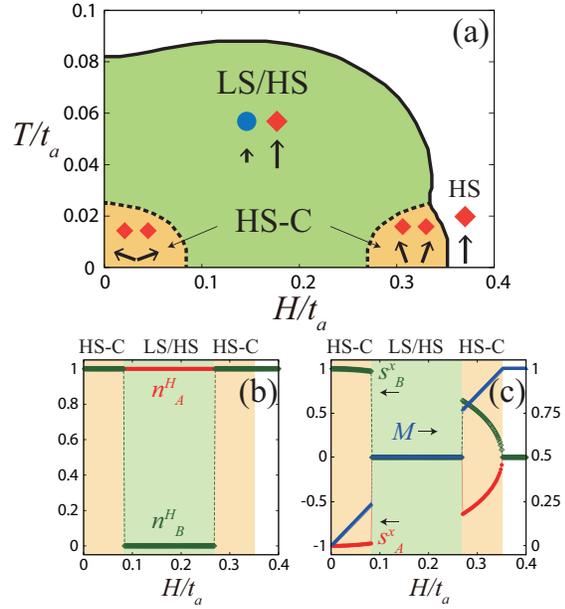}
\caption{(Color online) 
(a) Finite $T$ phase diagram under $H$ applied to the HS phase. 
Solid and broken lines represent the second- and first-order phase transitions, respectively. 
Circles, rhombuses, and their combined symbols represent the LS, HS, and EI states, respectively, 
and arrows represent schematic spin directions. 
Magnetic field dependences of (a) the HS densities ($n^H_{A, B}$), and (b) the $x$ components of the spin moments ($S^x_{A, B}$), and the magnetization ($M$) at $T=0$. 
Parameter values are chosen to be $J/t_a=4.6$, corresponding to a bold arrow in Fig.~\ref{fig:TJ}, and $t_b/t_a=-0.1$.  
}
\label{fig:MH2}
\end{figure}

The magnetic-field effect is also examined in the HS phase. 
In Fig.~\ref{fig:MH2}(a), we present the $H-T$ phase diagram at $J/t_a=4.6$ (see the bold arrow in Fig.~\ref{fig:TJ}). 
We find a reentrant transition for the HS phase as HS-C$\rightarrow$LS/HS$\rightarrow$HS-C with increasing $H$ at low $T$. 
Here, HS-C represents the HS phase with canted spin structures.   

Detailed $H$ dependences of the physical quantities are presented in Figs.~\ref{fig:MH2}(b) and \ref{fig:MH2}(c). 
The lower- and higher-critical magnetic fields, in which the phase transitions between the LS/HS and HS-C phases occur, are denoted as $H_1(\sim 0.08t_a)$ and $H_2(\sim 0.27t_a)$, respectively. 
Below $H_1$, the canted spin structure with a staggered $S^x$ alignment is induced by $H$ applied to the HS-AF phase. 
At $H_1$,  the HS states in the sublattice B are changed into the LS states, at which the spin moments are quenched, while the spins in the sublattice A are fully polarized. 
The Zeeman energy gain in this phase is larger than the canted AF structure where all sites are HS. 
This is owing to the fact that the exchange interaction between the HS sites in this phase does not work in the present model. 
The long-ranged exchange interaction between the HS sites may reduce the Zeeman energy gain, and shrink the LS/HS phase. 
A magnetization plateau with $M=1/2$ emerges in the $M-H$ curve.  
Above $H_2$, the energy gain owing to the ferromagnetic component in the HS-C phase overcomes that in the LS/HS phase, and the first-order phase transition occurs. 

\begin{figure}[t]
 \includegraphics[width=0.9\columnwidth,clip]{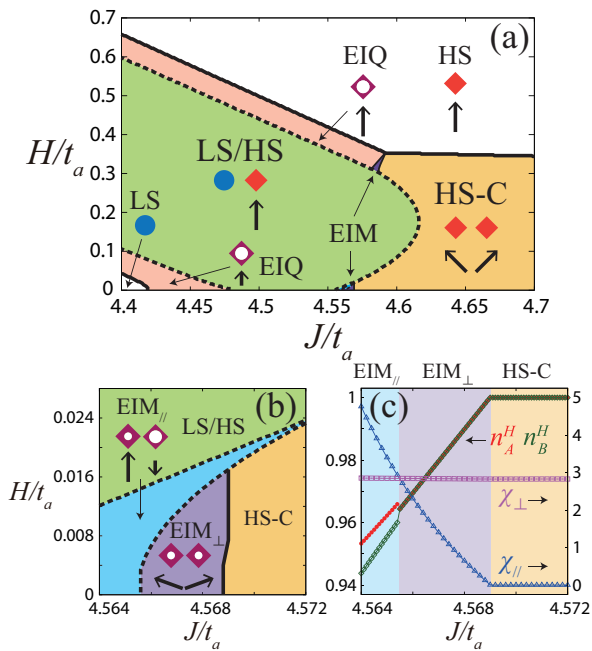}
\caption{(Color online) 
(a) Phase diagram in the $H-J$ plane at $T=0$. 
Solid and broken lines represent the second- and first-order phase transitions, respectively. 
Circles, rhombuses and their combined symbols represent the LS, HS and EI states, respectively, 
and arrows represent schematic spin directions.  
(b) An expansion of (a) around $4.564<J<4.572$ and $0<H<0.028$. 
Symbols EIM$_{\parallel}$ and EIM$_{\perp}$ indicate the EIM phase with the two SSDF sublattices, and that with a canted spin structure, respectively (see text). 
(c) The HS densities ($n^H_{A, B}$), and the transverse ($\chi_{\perp}$) and longitudinal ($\chi_{\parallel}$) magnetic susceptibilities. 
A parameter value is chosen to be $t_b/t_a=-0.1$. 
}
\label{fig:HJ}
\end{figure}
The $J-H$ phase diagram at $T=0$ is presented in Fig.~\ref{fig:HJ}(a). 
A rich variety of the $H$-induced phases are realized. 
It is confirmed that, with increasing $|t_b/t_a|$, the LS/HS phase shrinks and finally the two EIQ phases are merged into one (not shown in the figure).  
A characteristic $H$ effect appears around the EIM phase under the weak $H$ [see Fig.~\ref{fig:HJ}(b)]. 
With increasing $J$, a sequential phase transition occurs as 
LS/HS$\rightarrow$EIM$_{\parallel}$$\rightarrow$EIM$_{\perp}$$\rightarrow$HS-C in which the two kinds of EIM phases appear. 
In the EIM$_{\parallel}$ phase, the weights of LS and HS, i.e. $C_L$ and $C_H^\Gamma$ in Eq.~(\ref{eq:wfei}), are different in the two sublattices, and spins are directed along the $\pm z$ axis. 
On the other hand, in the EIM$_{\perp}$ phase, 
the wave functions are equivalent in the two sublattices, and spins are canted. 
These different spin structures are attributed to the anisotropy in the magnetic susceptibilities shown in Fig.~\ref{fig:HJ}(c), in which 
$\chi_{\perp}>\chi_{\parallel}$ ($\chi_{\perp}<\chi_{\parallel}$) in EIM$_{\parallel}$ (EIM$_{\perp}$) phase. This anisotropy originates from the following reasons.~\cite{nasu}
In the region of small $J$, the Zeeman energy is gained mainly with increasing $|C_H^\Gamma|$ in Eq.~(\ref{eq:wfei}). On the other hand, in the region of large $J$ where magnitudes of the magnetic moments are  enlarged, the spin canting is the main origin of the Zeeman energy gain.

Finally, relations between the present results and the recent experiments under the high magnetic field in LaCoO$_3$ are discussed. 
The first-order phase transition is observed around 60T below about 30-40K. 
Recent high magnetic field measurement up to about 120T reveals that additional first-order phase transition occurs between 80-100T in higher temperatures. 
Two kinds of the $H$-induced phases are confirmed: the low-$T$ and low-$H$ phase termed B1 and the high-$H$ phase termed B2. 
Present results provides a possible interpretation for the experimental observation.
Comparisons between the present results and the experiments are limited to semiquantitative discussions, since several factors, e.g. the intermediate-spin state, the lattice degree of freedom, and the magnetic anisotropy,  are not taken into account. 
We estimate that $H/t_a=0.025$ is about 100T, and $|t_b/t_a| \sim 0.2$ for the perovskite cobalt oxides. 
Thus, the phase diagrams shown in Figs.~\ref{fig:TH}(b) and (c) correspond to the experimental phase diagram.
There are some qualitative similarities between the theoretical $H-T$ phase diagram and the experimental one shown in Ref.~\citen{ikeda}. 
We propose a possible interpretation that the B1 and B2 phases are the LS/HS ordered phase and the EIQ phase, respectively. 
%

In summary, the magnetic field effects in a correlated electron system with SSDF are studied. 
The effective Hamiltonian derived from the two-orbital Hubbard model is analyzed by the MF approximation. 
We find that the magnetic field induces the EI and LS/HS phases, where 
the Zeeman energies are gained in the HS components. 
The sequential phase transitions occur in the LS phase as 
LS$\rightarrow$EIQ$\rightarrow$LS/HS$\rightarrow$EIQ$\rightarrow$HS, and 
in the HS phase as
HS-C$\rightarrow$LS/HS$\rightarrow$HS-C, where 
a reentrant phase transition for the HS phase occurs. 
The rich magnetic field effects are owing to the combinations of the LS and HS states in the wave function and in the real space configuration.  
The present results provide a possible interpretation for the recently observed $H$-induced phases in LaCoO$_3$. 
This study suggests that the correlated electron systems with SSDF is a quarry for novel quantum phenomena induced by the strong magnetic field, similar to the frustrated magnets.

We thank T.~Watanabe, A.~Ikeda, M.~Tokunaga, and Y.~Matsuda for their helpful discussions.
This work was supported by MEXT KAKENHI Grant No. 26287070 and 15H02100. 
Some of the numerical calculations were performed using the facilities of the Supercomputer Center, the Institute for Solid State Physics, the University of Tokyo.

\end{document}